\documentclass[
 aps,
 preprint,
 fleqn
]{revtex4}

\usepackage[T1]{fontenc}
\usepackage{amsmath}
\usepackage{amssymb}
\usepackage{amsfonts}

\usepackage{graphicx}

\usepackage{wick}

\newcommand{\psibar}[1]{\rlap{$\displaystyle \hspace{ #1 ex} \bar{\phantom{\psi}}$}}
\newcommand{\ME}[3]{\langle {#1} | {#2} | {#3} \rangle}

\newcommand{\pslash}{\rlap{$\displaystyle \hspace{-0.4ex} \diagup $}p}
\newcommand{\ppslash}{\rlap{$\displaystyle \hspace{-0.4ex} \diagup $}p'}

\newcommand{\Deltaslash}{\rlap{$\displaystyle \hspace{-0.1ex} \diagup $}\Delta}

\DeclareMathOperator{\intd}{d}
\DeclareMathOperator{\e}{e}
\DeclareMathOperator{\T}{T}
\DeclareMathOperator{\Tr}{Tr}

\begin{document}

\title{Leading Nucleon Form Factors in Single Gauge Boson Exchange Approximation}
\date{\today}
\author{Thorsten Sachs}
\author{Patrick Sturm}
\affiliation{Institut für Theoretische Physik II, Ruhr-Universität Bochum, D-44780 Bochum, Germany}

\begin{abstract}
We provide a complete calculation of all leading nucleon form factors at large momentum transfer and in single gauge boson exchange approximation. In order to evaluate the required nucleon transition probability matrix elements, we combine QCD perturbation theory with an expansion in nucleon distribution amplitudes. Using leading twist nucleon distribution amplitudes only, one obtains the desired nucleon form factors. The obtained results are consistent with experimental data when we use third order polynomials for the distribution amplitudes and non-perturbative models for the coefficients. Finally, we compare our results for the Dirac form factor of the proton with previous results based on the QCD factorization theorem, where a discrepancy about the correct symmetry factor is still under discussion. We examine the calculations and propose an explanation.
\end{abstract}

\maketitle

\section{Introduction}

Studies about leading nucleon form factors received much attention in the past. In the region of large momentum transfer, the formulation of the QCD factorization theorem in \cite{Lepage:1979a}, \cite{Lepage:1979b}, \cite{Lepage:1980}, and \cite{Efremov:1980a}, \cite{Efremov:1980b}, \cite{Efremov:1981} was a great success. In \cite{Lepage:1979b} and \cite{Lepage:1980}, the expressions for the leading electromagnetic form factors were derived. Using the QCD factorization theorem, these expressions were improved in \cite{Carlson:1987a}. Moreover, the representation for the isovector axial-vector form factor was created in \cite{Carlson:1986} with applications discussed in \cite{Carlson:1987c}. Similarly, the isoscalar axial-vector form factor was considered in \cite{Carlson:1987b}. Studying the behavior of vacuum to nucleon projection matrix elements, the leading electromagnetic form factors were calculated in \cite{Chernyak:1984a} and \cite{Chernyak:1984b} at the asymptotic limit. Hereby, an averaged value of $\alpha_s$ was used. This approach was also applied in \cite{Gari:1986} and improved in \cite{Stefanis:1989}. Furthermore, in \cite{Stefanis:1989}, also non-asymptotic contributions of the leading twist nucleon distribution amplitude were taken into account to calculate the leading electromagnetic form factors, see also \cite{Chernyak:1989}. Another approach is to keep $\alpha_s$ inside the integral, evaluated in \cite{Ji:1987}. Despite the success of perturbative QCD, the formalism was criticized in \cite{Isgur:1989}. Therefore, a modified formalism was discussed in \cite{Botts:1989}. Being more precise, the region of low momentum transfer was considered in the case of the pion, see \cite{Huang:1991}. Using the modified formalism in the case of the pion, the problems were cleared, see \cite{Li:1992}. This technique was also applied for the nucleon in \cite{Li:1993}. An improved calculation of the proton magnetic form factor, within the modified factorization scheme, was given in \cite{Bolz:1995a}. An analogous calculation for the neutron was done in \cite{Bolz:1995b}. A comprehensive work about the theory was published in \cite{Stefanis:1999}. A remaining problem has been pointed out in \cite{Brooks:2000}, where a different symmetry factor was used as in \cite{Ji:1987}. The same difference was also discussed in \cite{Thomson:2006}. Meanwhile,
nucleon distribution amplitudes of sub-leading twist were also extensively studied, see \cite{Braun:2000}, with applications in \cite{Braun:2006}.

This work is organized as follows. First, we will introduce the general construction for the desired form factors. Second, we will calculate the leading nucleon form factors. Next, we will compare our results with experimental data. Afterwards, we will discuss the normalization factor discrepancy. Finally, we draw our conclusions. Important technical details are collected in three appendices.

\section{General Construction of the Nucleon Form Factors}

Elastic lepton nucleon scattering in the single gauge boson exchange approximation is the basic process to receive information about the nucleon structure within QCD. Calculating the cross section of these processes, one has to study nucleon transition probability matrix elements including a single quark current. It is well known that they can be expressed by nucleon form factors. Expanding these matrix elements at large momentum transfer $Q^2=-q^2=-(P'-P)^2$, where $P$ and $P'$ are, respectively, the incoming and outgoing nucleon momenta, one can reduce the decomposition to one dominant term depending on the leading nucleon form factor. Overall, one gets four independent expressions.

It is noteworthy that the computation of the different form factors is similar from the theoretical perspective, whereas they differ in the experimental realization.

Nucleon matrix elements including the electromagnetic current can be expressed in terms of the the magnetic form factors or, equivalently, the Dirac form factors in the specified case, using obvious abbreviations,
\pagebreak[0]
\begin{eqnarray}
\ME{p(P')}{J_\mu^{em}(0)}{p(P)} &=& G_M^p(Q^2)\bar{N}(P')\gamma_\mu N(P) \\
\ME{n(P')}{J_\mu^{em}(0)}{n(P)} &=& G_M^n(Q^2)\bar{N}(P')\gamma_\mu N(P).
\end{eqnarray}

These structures describe the exchange of a highly virtual photon that is the dominant process in experiments. We notice that they are also part of neutral weak interactions.

In the the axial sector, we use the expansion in terms of the axial-vector form factors. In order to calculate them, it is necessary to choose the processes
\pagebreak[0]
\begin{eqnarray}
\ME{N_f(P')}{A_\mu^0(0)}{N_i(P)} &=& \frac{1}{2}G_A^s(Q^2)\bar{N}(P')\gamma_\mu\gamma_5\delta_{fi}N(P) \\
\ME{N_f(P')}{A_\mu^a(0)}{N_i(P)} &=& \frac{1}{2}G_A^v(Q^2)\bar{N}(P')\gamma_\mu\gamma_5\tau^a_{fi}N(P).
\end{eqnarray}

We work with two proton states for $G_A^s$, meaning that we have to evaluate the matrix element $\ME{p(P')}{\bar{\psi}_u(0)\gamma_\mu\gamma_5\psi_u(0)+\bar{\psi}_d(0)\gamma_\mu\gamma_5\psi_d(0)}{p(P)}=G_A^s(Q^2)\bar{N}(P')\gamma_\mu\gamma_5N(P)$. Concerning $G_A^v$, we consider the proton to neutron transition. We end up with the reduced matrix element $\ME{n(P')}{\bar{\psi}_d(0)\gamma_\mu\gamma_5\psi_u(0)}{p(P)}=G_A^v(Q^2)\bar{N}(P')\gamma_\mu\gamma_5N(P)$.

The isovector current matrix elements describe a part of the charged weak interactions generated by a $W^+$ or $W^-$ boson and the neutral weak interactions mediated by a $Z^0$ boson. The isoscalar current matrix elements only appear in non standard weak interaction theories, such as the exchange of an extra $Z^0$ boson.

\section{Calculation of the Leading Nucleon Form Factors}

Starting from the given four matrix elements, we make use of the large momentum transfer for a perturbative expansion. Therefore, we need the S-Matrix containing the interaction part of the QCD Lagrangian. In all cases, one has to expand up to the fourth order to obtain non-vanishing results. We mentioned that the calculations of the different form factors are similar, so we demonstrate the evaluation of the most prominent member, the proton magnetic form factor
\pagebreak[0]
\begin{equation}
\label{e}
\frac{(4\pi\bar{\alpha}_s)^2}{24}\ME{p(P')}{\sum_qe_q\bar{\psi}_q(0)\gamma_\mu\psi_q(0)
\T\left[\prod_{i=1}^4\int\intd^4\!x_i\sum_{q_i}\bar{\psi}_{q_i}(x_i)\gamma_{\alpha_i}A^{\alpha_i}(x_i)\psi_{q_i}(x_i)\right]}{p(P)}.
\end{equation}

This expansion can be expressed by Wick Contractions and Feynman Diagrams. These diagrams have similar structures and therefore, we will calculate just one diagram in detail. The entire diagram classification will be specified in the appendices.

\subsection{Sample Diagram Evaluation}

The chosen diagram can be presented as follows. We have the incoming proton on the left and we have the outgoing proton on the right. The cross denotes the coupling to the virtual photon and the quark lines denote $u$, $u$, $d$ from top to bottom. The designations at the vertices are the corresponding coordinates and the designations at the lines are the corresponding momenta.

\begin{center}
\includegraphics[scale=1.00]{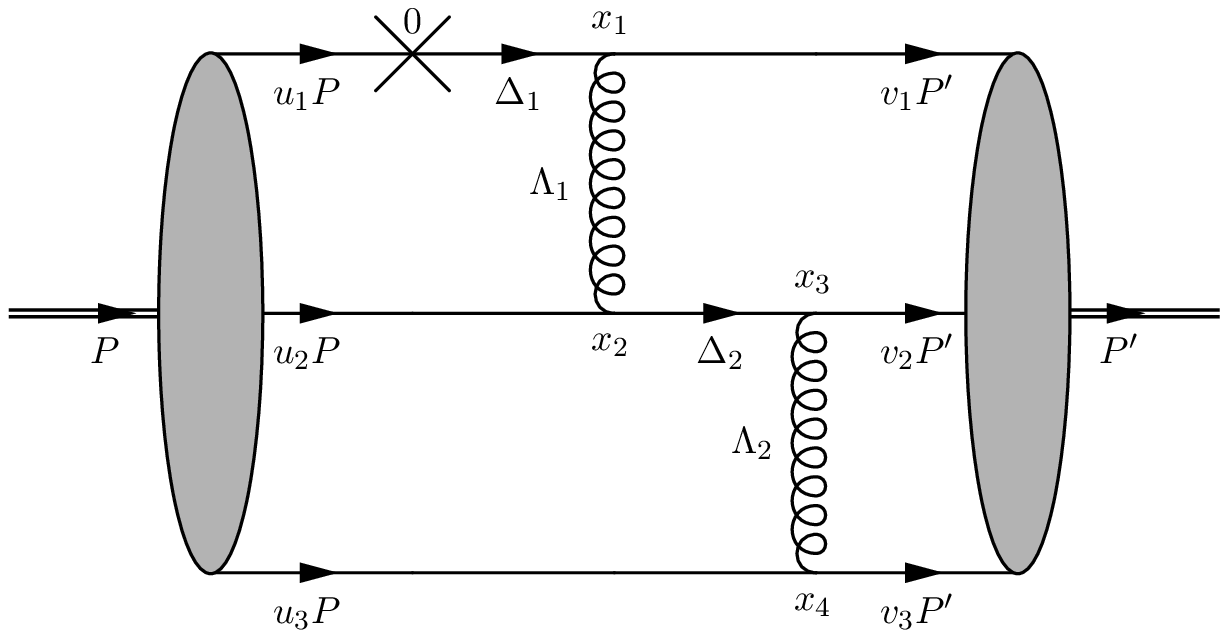}
\end{center}

Let us begin with the determination of the color factor. Therefore, one has to examine the color structure of the diagram, denoting the color indices with ($a,\ldots,i$),
\pagebreak[0]
\begin{equation}
\begin{split}
& \psibar{9.9} \wick[u]{1}{[<1\psi_u(x_1)]_c[>1\psi_u(0)]_a}
\psibar{10.0} \wick[u]{1}{[<1\psi_u(x_3)]_g[>1\psi_u(x_2)]_d}
\wick[u]{1}{[t^{a_1}]_{bc}[t^{a_2}]_{de}<1A_{\alpha_1}^{a_1}(x_1)>1A_{\alpha_2}^{a_2}(x_2)}
\wick[u]{1}{[t^{a_3}]_{fg}[t^{a_4}]_{hi}<1A_{\alpha_3}^{a_3}(x_3)>1A_{\alpha_4}^{a_4}(x_4)} \\
& \ME{p(P')}{[\bar{\psi}_u(x_1)]_b[\bar{\psi}_u(x_3)]_f[\bar{\psi}_d(x_4)]_h}{0}\ME{0}{[\psi_u(0)]_a[\psi_u(x_2)]_e[\psi_d(x_4)]_i}{p(P)}.
\end{split}
\nonumber
\end{equation}
Combining all terms and contracting the generators, one gets the result of the color factor
\pagebreak[0]
\begin{equation}
\label{c}
\mathcal{C}_F = \frac{1}{6}\varepsilon_{bfh}\varepsilon_{aei}\delta_{ca}\delta_{gd}[t^{a_1}]_{bc}[t^{a_2}]_{de}[t^{a_3}]_{fg}[t^{a_4}]_{hi}
\delta^{a_1a_2}\delta^{a_3a_4} = \frac{4}{9}.
\end{equation}

This result is the same for all diagrams of $G_M^p$, $G_M^n$ and $G_A^s$. The flavor change in the diagrams of $G_A^v$ leads to the exchange of
$\varepsilon_{bfh} \rightarrow \varepsilon_{bhf}$ producing an overall minus sign.

We continue with the evaluation of the Lorentz structure of the diagram, designating the Lorentz indices by ($a,\ldots,j$). One gets an expression as part of (\ref{e}), where we already included (\ref{c}), so that
\pagebreak[0]
\begin{equation}
\begin{split}
& -\frac{(4\pi\bar{\alpha}_s)^2}{54}e_u\prod_{i=1}^4\int\intd^4\!x_i
[\gamma_\mu]_{ab}[\gamma_{\alpha_1}]_{cd}[\gamma_{\alpha_2}]_{ef}[\gamma_{\alpha_3}]_{gh}[\gamma_{\alpha_4}]_{ij} \\
& \psibar{10.0}
\wick[u]{1}{[<1\psi_u(x_1)]_d[>1\psi_u(0)]_a}
\psibar{10.1}
\wick[u]{1}{[<1\psi_u(x_3)]_h[>1\psi_u(x_2)]_e}
\wick[u]{1}{<1A^{\alpha_1}(x_1)>1A^{\alpha_2}(x_2)}
\wick[u]{1}{<1A^{\alpha_3}(x_3)>1A^{\alpha_4}(x_4)} \\
& \ME{p(P')}{[\bar{\psi}_u(x_1)]_c[\bar{\psi}_u(x_3)]_g[\bar{\psi}_d(x_4)]_i}{0}
\ME{0}{[\psi_u(0)]_b[\psi_u(x_2)]_f[\psi_d(x_4)]_j}{p(P)}.
\end{split}
\nonumber
\end{equation}

The formulas for the propagators are standard and the representations of the vacuum to nucleon projection matrix elements are extensively studied in \cite{Braun:2000}. Applications are discussed in \cite{Braun:2006}. Hereby, $g^{\alpha_i\alpha_j}$ is the metric tensor. Recombining the Lorentz structures, one gets the following result
\pagebreak[0]
\begin{equation}
\begin{split}
& \frac{(4\pi\bar{\alpha}_s)^2}{864}e_u
\prod_{i=1}^4\int\frac{\intd^4\!x_i}{(2\pi)^4}\prod_{j=1}^2\int\frac{\intd^4\!\Delta_j}{\Delta_j^2+i0}\prod_{k=1}^2\int\frac{\intd^4\!\Lambda_k}{\Lambda_k^2+i0}
\int[\intd\!u][\intd\!v]g^{\alpha_1\alpha_2}g^{\alpha_3\alpha_4}\mathcal{S} \\
& \e^{-ix_1\cdot(\Delta_1-\Lambda_1-v_1p')}\e^{-ix_2\cdot(-\Delta_2+\Lambda_1+u_2p)}\e^{-ix_3\cdot(\Delta_2-\Lambda_2-v_2p')}\e^{-ix_4\cdot(\Lambda_2+u_3p-v_3p')}.
\end{split}
\nonumber
\end{equation}

The integration over the coordinates produce delta functions which can be used to integrate over the momenta. They describe the momentum conservation at each vertex of the diagram
\pagebreak[0]
\begin{equation}
\begin{array}{lll}
\Delta_1 = (v_1+v_2+v_3)p'-(u_2+u_3)p & \phantom{v_2p'}\qquad & \Delta_2 = (v_2+v_3)p'-u_3p \\
\Lambda_1 = (v_2+v_3)p'-(u_2+u_3)p & & \Lambda_2 = v_3p'-u_3p.
\end{array}
\nonumber
\end{equation}

The undeclared component $\mathcal{S}$ is the sum of all remaining structures connected with combinations of nucleon distribution amplitudes, where $V$, $A$, and $T$ denote vector, axial-vector, and tensor structures, respectively, and nucleon spinors. For convenience, we omit the twist index at the distribution amplitudes and the dependence on the quark momentum fractions. Moreover, we only have to deal with the large component of the spinor, so we can use the standard notation.  One gets
\pagebreak[0]
\begin{eqnarray}
\mathcal{S}_1 &=& \bar{N}(P')\gamma_{\alpha_4}N(P)
\Tr [\gamma_\mu\pslash\gamma_{\alpha_2}\Deltaslash_2\gamma_{\alpha_3}\ppslash\gamma_{\alpha_1}\Deltaslash_1](VV+AA) \nonumber \\
\mathcal{S}_2 &=& \bar{N}(P')\gamma_{\alpha_4}\gamma_5N(P)
\Tr [\gamma_\mu\pslash\gamma_{\alpha_2}\Deltaslash_2\gamma_{\alpha_3}\gamma_5\ppslash\gamma_{\alpha_1}\Deltaslash_1](-AV-VA) \nonumber \\
\mathcal{S}_3 &=& \bar{N}(P')\gamma^{\lambda'}\gamma_{\alpha_4}\gamma^{\lambda}N(P)
\Tr [\gamma_\mu i\sigma_{\lambda p}\gamma_{\alpha_2}\Deltaslash_2\gamma_{\alpha_3}i\sigma_{\lambda' p'}\gamma_{\alpha_1}\Deltaslash_1](-TT). \nonumber
\end{eqnarray}

Finally, we can express the result of the discussed diagram depending on the integration over the quark momentum fractions in the form
\pagebreak[0]
\begin{equation}
\frac{(4\pi\bar{\alpha}_s)^2}{216}\frac{e_u}{Q^4}\int\frac{[\intd\!u]}{u_3(u_2+u_3)^2}\frac{[\intd\!v]}{v_3(v_2+v_3)^2}\left[(V-A)^2+4T^2\right]
\bar{N}(P')\gamma_\mu N(P).
\end{equation}

\subsection{Results for the Form Factors}

We add the obtained results of the contributing diagrams for every form factor and insert the polynomial expression of the independent leading twist nucleon distribution amplitude, see \cite{Stefanis:1999} and \cite{Braun:2000}. As already explained, we apply third order polynomials expressible as $\Phi_3(u_1,u_2,u_3)=120u_1u_2u_3f_N[c_1+c_2u_1+c_3u_3+c_4u_1^2+c_5u_3^2+c_6u_1u_3]$. The required coefficients are non-perturbative parameters which are scale dependent and have to be modeled or taken from the lattice. We use the standard hadronic scale of $1\,\textnormal{GeV}$, and adopt
$f_N=(5.3\pm 0.5)\cdot 10^{-3}\,\textnormal{GeV}^2$, see \cite{Braun:2000}. Meanwhile, this parameter is also studied on the lattice, see \cite{Braun:2009}. The first and second order coefficients are taken from \cite{Braun:2000} and the third order modeled coefficients can be found in \cite{Stefanis:1999}. We use the models of Chernyak, Zhitnitsky (CZ) \cite{Chernyak:1984a}, Gari, Stefanis (GS) \cite{Gari:1986}, King, Sachrajda (KS) \cite{King:1987}, Chernyak, Ogloblin, Zhitnitsky (COZ) \cite{Chernyak:1989}, and Stefanis, Bergmann (HET) \cite{Stefanis:1993}. The abbreviation HET means heterotic conception.

Every form factor can be expressed by a corresponding coefficient function as $Q^4G(Q^2)=(4\pi\bar{\alpha}_s)^2f_N^2X(c_1,c_2,c_3,c_4,c_5,c_6)$. Finally, we have the following list of them:

\begin{equation}
X_M^p(c_1,c_2,c_3,c_4,c_5,c_6)=\frac{25}{11664}\sum_{i=1}^{6}\sum_{j=1}^{i}a_{ij}c_ic_j
\end{equation}
with $a_{11}=0$, $a_{21}=4320$, $a_{31}=-1728$, $a_{41}=6048$, $a_{51}=432$, $a_{61}=-1296$, $a_{22}=2160$, $a_{32}=672$, $a_{42}=4536$, $a_{52}=1152$, $a_{62}=-240$, $a_{33}=-528$, $a_{43}=1464$, $a_{53}=-168$, $a_{63}=-480$, $a_{44}=2123$, $a_{54}=1418$, $a_{64}=72$, $a_{55}=95$, $a_{65}=-180$, $a_{66}=-72$,

\begin{equation}
X_M^n(c_1,c_2,c_3,c_4,c_5,c_6)=\frac{25}{5832}\sum_{i=1}^{6}\sum_{j=1}^{i}a_{ij}c_ic_j
\end{equation}
with $a_{11}=3888$, $a_{21}=432$, $a_{31}=3456$, $a_{41}=-972$, $a_{51}=1836$, $a_{61}=648$, $a_{22}=-336$, $a_{32}=144$, $a_{42}=-960$, $a_{52}=-180$, $a_{62}=120$, $a_{33}=1008$, $a_{43}=-336$, $a_{53}=1392$, $a_{63}=240$, $a_{44}=-491$, $a_{54}=-395$, $a_{64}=0$, $a_{55}=523$, $a_{65}=126$, $a_{66}=24$,

\begin{equation}
X_A^v(c_1,c_2,c_3,c_4,c_5,c_6)=\frac{25}{1944}\sum_{i=1}^{6}\sum_{j=1}^{i}a_{ij}c_ic_j
\end{equation}
with $a_{11}=1296$, $a_{21}=1872$, $a_{31}=144$, $a_{41}=1872$, $a_{51}=216$, $a_{61}=-144$, $a_{22}=640$, $a_{32}=448$, $a_{42}=1220$, $a_{52}=432$, $a_{62}=-8$, $a_{33}=-176$, $a_{43}=512$, $a_{53}=-252$, $a_{63}=-72$, $a_{44}=549$, $a_{54}=440$, $a_{64}=42$, $a_{55}=-100$, $a_{65}=-30$, $a_{66}=-12$,

\begin{equation}
X_A^s(c_1,c_2,c_3,c_4,c_5,c_6)=-\frac{25}{3888}\sum_{i=1}^{6}\sum_{j=1}^{i}a_{ij}c_ic_j
\end{equation}
with $a_{11}=7776$, $a_{21}=3456$, $a_{31}=3456$, $a_{41}=-216$, $a_{51}=-216$, $a_{61}=1728$, $a_{22}=-304$, $a_{32}=1472$, $a_{42}=-1560$, $a_{52}=312$, $a_{62}=480$, $a_{33}=-304$, $a_{43}=312$, $a_{53}=-1560$, $a_{63}=480$, $a_{44}=-907$, $a_{54}=-124$, $a_{64}=144$, $a_{55}=-907$, $a_{65}=144$, $a_{66}=72$.

\section{Comparison with Experimental Data}

The final step is to compare the results for the different form factors with available experimental data. In order to compare our results with those existing in the literature, we do not specify the values of the parameters $f_N$ and $\alpha_s$.

The determination of $\alpha_s$ has to be done with care, because it depends on the momentum transfer of the process, or more precisely, it depends on the gluon virtualities. Consequently, one gets two of them for every diagram independently. The main problem arises from the soft gluon region, where $\alpha_s$ becomes divergent. It is possible to apply different estimations to avoid this problem. In principle, one can introduce an averaged $\bar{\alpha}_s$ for all diagrams, see \cite{Chernyak:1984a}, \cite{Chernyak:1984b}. This approach was also applied in \cite{Gari:1986} and improved in \cite{Stefanis:1989} for each diagram separately. Hereby, the value of $\bar{\alpha}_s\approx 0.3$ was used. Another approach is to keep $\alpha_s$ inside the integral, as done in \cite{Ji:1987}. This technique can change the slope of the form factors and improve the scaling behavior of them. Nevertheless, this cannot be done without introducing a new cutoff parameter. Taking into account soft gluon corrections \cite{Li:1993}, one may claim that the area of soft gluon virtualities is systematically underestimated. According to this, we prefer to use an averaged value of $\bar{\alpha}_s$ which is taken from an effective momentum transfer around $1\,\textnormal{GeV}^2$. Consequently, we use $\bar{\alpha}_s=0.45\pm 0.05$. One should note that the obtained results contain large uncertainties due to the choice of $\alpha_s$.

The first order polynomial describes the asymptotic limit of $Q^2$ leading to the following results
$Q^4G_M^p(Q^2)=0(4\pi\bar{\alpha}_s)^2f_N^2=0\,\textnormal{GeV}^4$, and
$Q^4G_M^n(Q^2)=50/3(4\pi\bar{\alpha}_s)^2f_N^2=(15\pm 5)\cdot 10^{-3}\,\textnormal{GeV}^4$, and
$Q^4G_A^v(Q^2)=50/3(4\pi\bar{\alpha}_s)^2f_N^2=(15\pm 5)\cdot 10^{-3}\,\textnormal{GeV}^4$, and
$Q^4G_A^s(Q^2)=-50(4\pi\bar{\alpha}_s)^2f_N^2=-(45\pm 5)\cdot 10^{-3}\,\textnormal{GeV}^4$.

The second order polynomial is still related to values of $Q^2$ which are unreachable in experiments, but one obtains a non-asymptotic behavior. We get the results
$Q^4G_M^p(Q^2)=166(4\pi\bar{\alpha}_s)^2f_N^2=(0.15\pm 0.05)\,\textnormal{GeV}^4$, and
$Q^4G_M^n(Q^2)=-74(4\pi\bar{\alpha}_s)^2f_N^2=-(0.07\pm 0.05)\,\textnormal{GeV}^4$, and
$Q^4G_A^v(Q^2)=247(4\pi\bar{\alpha}_s)^2f_N^2=(0.22\pm 0.05)\,\textnormal{GeV}^4$, and
$Q^4G_A^s(Q^2)=256(4\pi\bar{\alpha}_s)^2f_N^2=(0.23\pm 0.05)\,\textnormal{GeV}^4$.

The third order polynomial can be correlated to values of $Q^2$ in the area around $10\,\textnormal{GeV}^2$. In this case, one can compare with experimental data, though one should keep in mind that the soft contribution may still be sizeable, see \cite{Bolz:1995a} and \cite{Bolz:1995b}.

For the proton magnetic form factor we obtain
\pagebreak[0]
\begin{equation}
Q^4G_M^p(Q^2)=(4\pi\bar{\alpha}_s)^2f_N^2\cdot
\left\{
\begin{array}{c}
1.16\cdot 10^3 \\
1.16\cdot 10^3 \\
1.69\cdot 10^3 \\
1.34\cdot 10^3 \\
1.53\cdot 10^3
\end{array}
\right\}
=
\left\{
\begin{array}{cc}
(1.0\pm 0.1)\enspace\textnormal{GeV}^4 & \textnormal{(CZ)} \\
(1.0\pm 0.1)\enspace\textnormal{GeV}^4 & \textnormal{(GS)} \\
(1.5\pm 0.1)\enspace\textnormal{GeV}^4 & \textnormal{(KS)} \\
(1.2\pm 0.1)\enspace\textnormal{GeV}^4 & \textnormal{(COZ)} \\
(1.4\pm 0.1)\enspace\textnormal{GeV}^4 & \textnormal{(HET)}.
\end{array}
\right.
\end{equation}
The electromagnetic interaction ensures that this form factor can be measured at large $Q^2$. Moreover, the proton is a stable particle and so the measurement can directly be done on the proton. We have data of $G_M^p$ for a comprehensive region, see \cite{Arnold:1986} and \cite{Sill:1993}. The obtained results are in sufficient agreement with the averaged experimental value $Q^4G_M^p(Q^2)=(1.0\pm 0.1)\,\textnormal{GeV}^4$.

For the neutron magnetic form factor we find
\pagebreak[0]
\begin{equation}
Q^4G_M^n(Q^2)=(4\pi\bar{\alpha}_s)^2f_N^2\cdot
\left\{
\begin{array}{c}
-0.56\cdot 10^3 \\
-0.11\cdot 10^3 \\
-0.70\cdot 10^3 \\
-0.63\cdot 10^3 \\
-0.16\cdot 10^3
\end{array}
\right\}
=
\left\{
\begin{array}{cc}
-(0.5\pm 0.1)\enspace\textnormal{GeV}^4 & \textnormal{(CZ)} \\
-(0.1\pm 0.1)\enspace\textnormal{GeV}^4 & \textnormal{(GS)} \\
-(0.6\pm 0.1)\enspace\textnormal{GeV}^4 & \textnormal{(KS)} \\
-(0.6\pm 0.1)\enspace\textnormal{GeV}^4 & \textnormal{(COZ)} \\
-(0.1\pm 0.1)\enspace\textnormal{GeV}^4 & \textnormal{(HET)}.
\end{array}
\right.
\end{equation}
The electromagnetic interaction ensures that this form factor can also be measured at large $Q^2$. Unfortunately, the neutron is not a stable particle and so the measurement cannot be done on the neutron directly. Data of $G_M^n$ in the lower region of $Q^2$, see \cite{Rock:1982} and \cite{Rock:1992}, are available, but an extrapolation to the upper area is also possible. The obtained results are in sufficient agreement with the extrapolated and averaged experimental value $Q^4G_M^n(Q^2)=-(0.5\pm 0.1)\,\textnormal{GeV}^4$.

The next result concerns the isovector axial-vector form factor
\pagebreak[0]
\begin{equation}
Q^4G_A^v(Q^2)=(4\pi\bar{\alpha}_s)^2f_N^2\cdot
\left\{
\begin{array}{c}
1.75\cdot 10^3 \\
1.31\cdot 10^3 \\
2.50\cdot 10^3 \\
2.04\cdot 10^3 \\
1.80\cdot 10^3
\end{array}
\right\}
=
\left\{
\begin{array}{cc}
(1.6\pm 0.1)\enspace\textnormal{GeV}^4 & \textnormal{(CZ)} \\
(1.2\pm 0.1)\enspace\textnormal{GeV}^4 & \textnormal{(GS)} \\
(2.2\pm 0.1)\enspace\textnormal{GeV}^4 & \textnormal{(KS)} \\
(1.8\pm 0.1)\enspace\textnormal{GeV}^4 & \textnormal{(COZ)} \\
(1.6\pm 0.1)\enspace\textnormal{GeV}^4 & \textnormal{(HET)}.
\end{array}
\right.
\end{equation}
This form factor appears in standard weak interactions. The consequence is that the desired processes are generally suppressed by the dominant electromagnetic interaction, but it is possible to choose processes for the measurement which can be separated from the electromagnetic interaction. Avoiding the electromagnetic domination, one has to apply neutrino neutron scattering. The neutrino interacts slightly and so we have data for values of $Q^2$ which are less than required, see \cite{Kitagaki:1983}. Predicting the behavior in the required region, one can extract a value for $G_A^v$ which is given by
$Q^4G_A^v(Q^2)=(1.5\pm 0.2)\,\textnormal{GeV}^4$.

Last, we consider the isoscalar axial-vector form factor
\pagebreak[0]
\begin{equation}
Q^4G_A^s(Q^2)=(4\pi\bar{\alpha}_s)^2f_N^2\cdot
\left\{
\begin{array}{c}
1.66\cdot 10^3 \\
3.10\cdot 10^3 \\
2.61\cdot 10^3 \\
1.89\cdot 10^3 \\
3.71\cdot 10^3
\end{array}
\right\}
=
\left\{
\begin{array}{cc}
(1.5\pm 0.1)\enspace\textnormal{GeV}^4 & \textnormal{(CZ)} \\
(2.8\pm 0.1)\enspace\textnormal{GeV}^4 & \textnormal{(GS)} \\
(2.3\pm 0.1)\enspace\textnormal{GeV}^4 & \textnormal{(KS)} \\
(1.7\pm 0.1)\enspace\textnormal{GeV}^4 & \textnormal{(COZ)} \\
(3.3\pm 0.1)\enspace\textnormal{GeV}^4 & \textnormal{(HET)}.
\end{array}
\right.
\end{equation}
This form factor appears in non standard weak interaction theories only. Consequently, the required processes are suppressed by all discussed interaction theories. Therefore, it is not possible to construct an experiment in order to measure this form factor independently. We have experimental data on $G_A^s$ for values of $Q^2$ which are significantly less than required. Under these circumstances, it makes no sense to compare the results. According to this, we can only consider our results as a prediction for further experiments.

\section{Discussion of the Normalization Factor Discrepancy}

Finally, we will compare our results for the proton magnetic form factor with previous results obtained in calculations based on the QCD factorization theorem, see \cite{Lepage:1979a}, \cite{Lepage:1979b}, \cite{Lepage:1980}, and \cite{Efremov:1980a}, \cite{Efremov:1980b}, \cite{Efremov:1981}. Hereby, we focus on the known discrepancy concerning the symmetry factor. The starting point of our considerations is the calculation of the desired form factor performed in \cite{Ji:1987}. After some time, the discussed form factor was recalculated in \cite{Brooks:2000}. They obtained a result which is precisely a factor of $2$ smaller than the previous result. They mentioned that they do not understand the origin of this discrepancy. A newer calculation of the proton magnetic form factor was done in \cite{Thomson:2006}. Their result is in agreement with \cite{Brooks:2000}. They realized that the symmetry factor in \cite{Ji:1987} is precisely a factor of $\sqrt{2}$ larger than in \cite{Brooks:2000}. Nevertheless, they mentioned that they do not know which symmetry factor is correct. However, our result of the proton magnetic form factor is the same as in \cite{Thomson:2006}. At this point, we should try to find the origin of the discrepancy between the older and the newer results. Looking in \cite{Lepage:1980}, we can see that the normalization of the soft parts is not fixed, but a required normalization connected to the number of colors is discussed. We realized that we need a normalization connected to the number of possible color states. In the case of the pion, one gets no difference, because the normalization by the number of colors as well as the normalization by the number of possible color states would lead to the factor $1/\sqrt{3}$. This is not true in the case of the nucleon. The normalization by the number of colors would obviously lead to the factor $1/\sqrt{3}$ too, but the normalization by the number of possible color states would require the factor $1/\sqrt{6}$ creating a difference of $\sqrt{2}$. Because two soft parts are required, one has to square the normalization factor. We claim that this is the source for the factor of $2$ difference in previous calculations. Indeed, when we compare the normalization of \cite{Ji:1987} and \cite{Brooks:2000}, we find the normalization factor $1/\sqrt{3}$ in \cite{Ji:1987} and $1/\sqrt{6}$ in \cite{Brooks:2000} as expected.

\section{Conclusion}

In this work, we calculated all leading nucleon form factors at large momentum transfer and in single gauge boson exchange approximation.

In order to compare the obtained results with experimental data, we used third order polynomials for the distribution amplitudes and non-perturbative models for the coefficients.

The calculated results for the form factors are essentially in good agreement with the available experimental data. Nevertheless, further experimental data at large momentum transfer are required in order to gain a deeper understanding of the nucleon form factors.

Finally, we discussed the known discrepancy concerning the symmetry factor in previous works. We argued that taking into account the color normalization appropriately, the discrepancy can be resolved.

\appendix

\section{Diagrams}

The general decomposition in diagrams can be classified by the quark coupling to the specified current in the corresponding matrix element. In principle, one gets 14 diagrams for every quark. The only exception is that the transition expression of $G_A^v$ does not allow the coupling to the quark which occurs once in the initial nucleon. The discussed diagrams are not independent, because two quarks in the initial nucleon are identical and a general interchange of the incoming and outgoing nucleon does not change the final result.

Fixing the positions of the quark lines in the manner that the single quark in the initial nucleon is at the top or bottom position, one has to deal with 7 diagrams as follows.

\begin{center}
\includegraphics[scale=0.22]{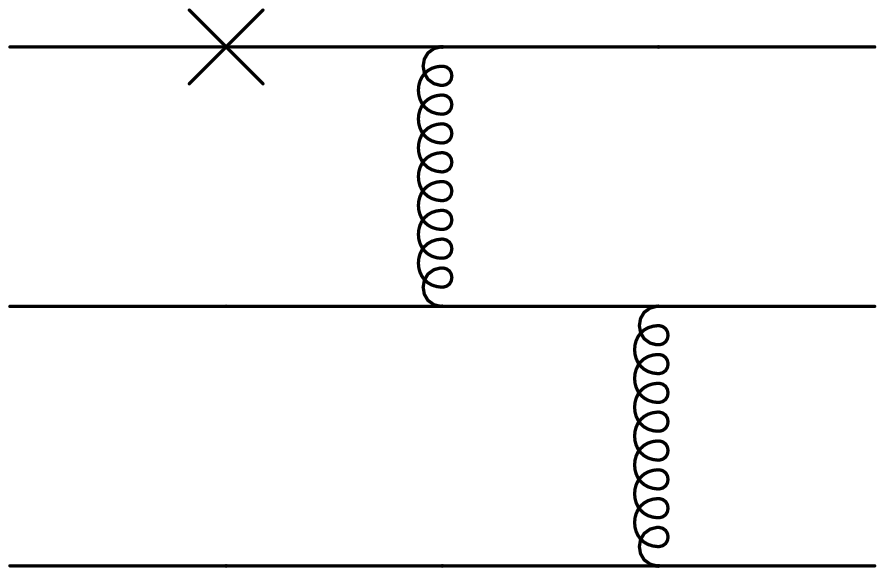}
\includegraphics[scale=0.22]{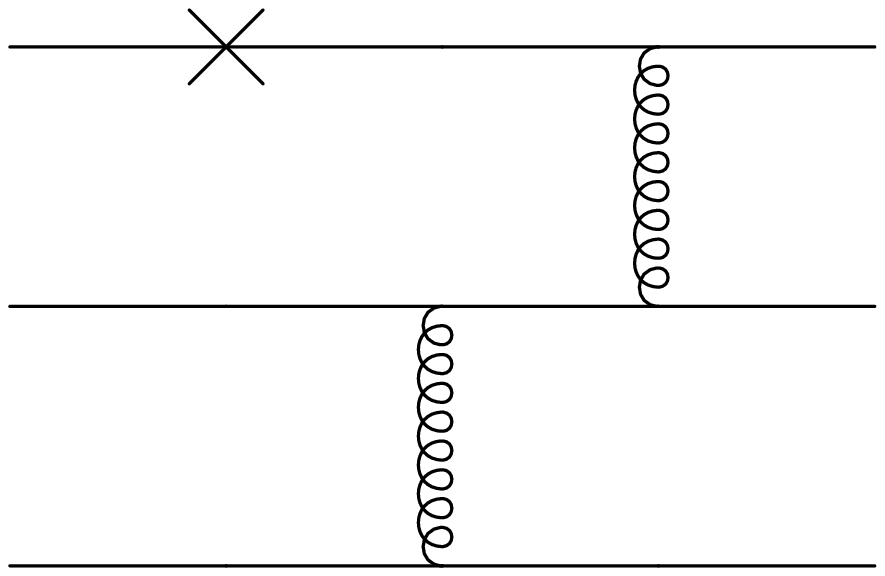}
\includegraphics[scale=0.22]{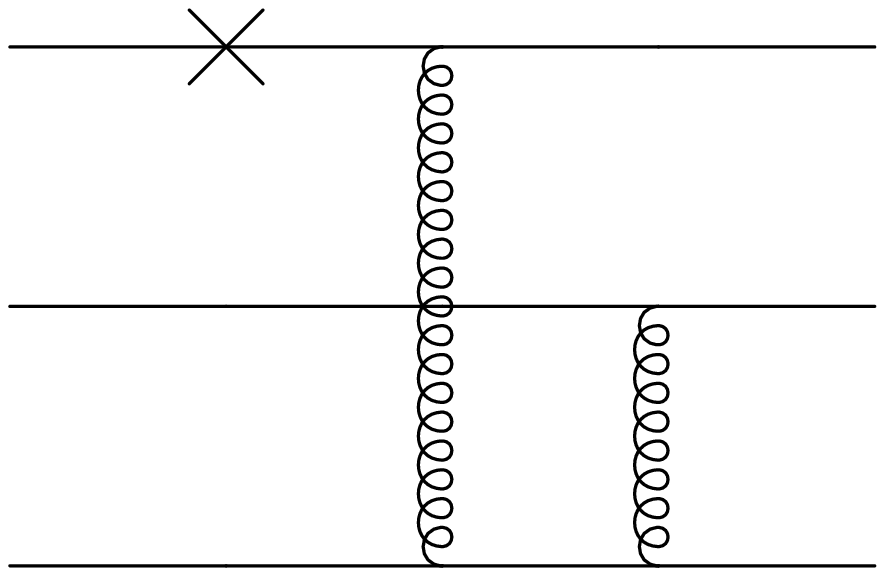}
\includegraphics[scale=0.22]{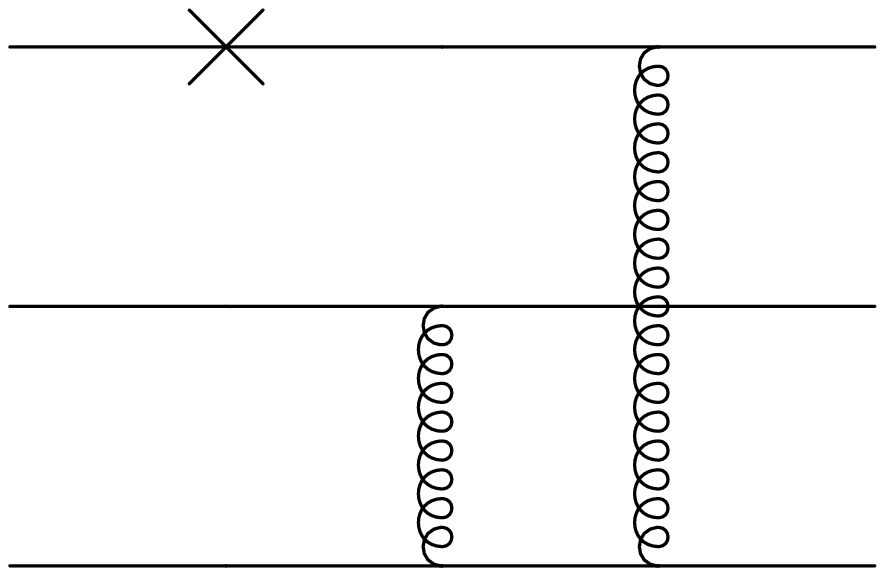}
\includegraphics[scale=0.22]{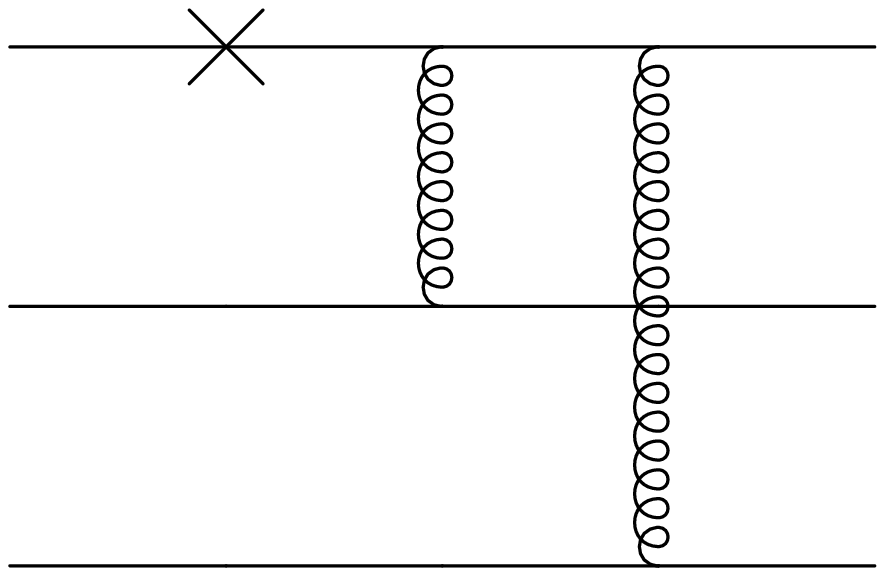}
\includegraphics[scale=0.22]{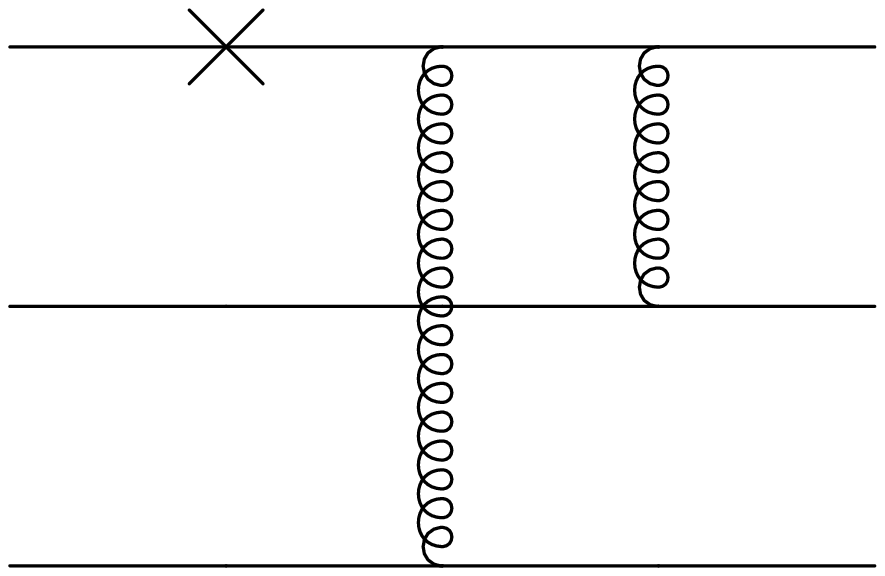}
\includegraphics[scale=0.22]{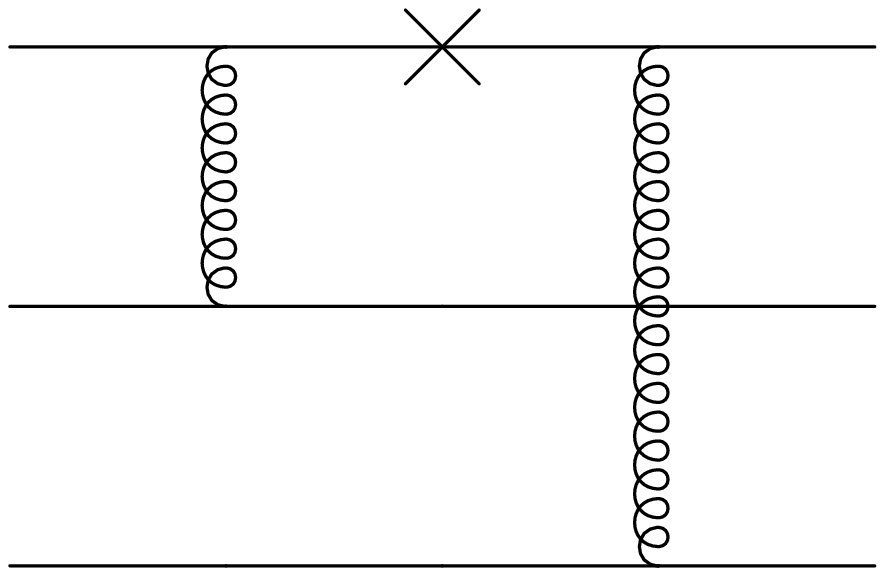}
\end{center}

We can derive the following list of momentum conservation constraints when the single quark in the initial nucleon is at the bottom position. The expressions for the diagrams with the considered quark at the top position can be obtained by the interchange of $u_1 \leftrightarrow u_3$ and $v_1 \leftrightarrow v_3$. This list can be used for all form factors except for $G_A^v$ where an exchange of $v_2 \leftrightarrow v_3$ is required
\pagebreak[0]
\begin{equation}
\begin{array}{lll}
\Delta_1 = (v_1+v_2+v_3)p'-(u_2+u_3)p & \phantom{v_2p'}\qquad & \Delta_2 = (v_2+v_3)p'-u_3p \\
\Lambda_1 = (v_2+v_3)p'-(u_2+u_3)p & & \Lambda_2 = v_3p'-u_3p
\end{array}
\nonumber
\end{equation}
\begin{equation}
\begin{array}{lll}
\Delta_1 = (v_1+v_2+v_3)p'-(u_2+u_3)p & \phantom{v_2p'}\qquad & \Delta_2 = (u_2+u_3)p-v_3p' \\
\Lambda_1 = (v_2+v_3)p'-(u_2+u_3)p & & \Lambda_2 = v_3p'-u_3p
\end{array}
\nonumber
\end{equation}
\begin{equation}
\begin{array}{lll}
\Delta_1 = (v_1+v_2+v_3)p'-(u_2+u_3)p & \phantom{v_2p'}\qquad & \Delta_2 = (v_2+v_3)p'-u_2p \\
\Lambda_1 = (v_2+v_3)p'-(u_2+u_3)p & & \Lambda_2 = u_2p-v_2p'
\end{array}
\nonumber
\end{equation}
\begin{equation}
\begin{array}{lll}
\Delta_1 = (v_1+v_2+v_3)p'-(u_2+u_3)p & \phantom{v_2p'}\qquad & \Delta_2 = (u_2+u_3)p-v_2p' \\
\Lambda_1 = (v_2+v_3)p'-(u_2+u_3)p & & \Lambda_2 = u_2p-v_2p'
\end{array}
\nonumber
\end{equation}
\begin{equation}
\begin{array}{lll}
\Delta_1 = (v_1+v_2+v_3)p'-(u_2+u_3)p & \phantom{v_2p'}\qquad & \Delta_2 = (v_1+v_3)p'-u_3p \\
\Lambda_1 = v_2p'-u_2p & & \Lambda_2 = v_3p'-u_3p
\end{array}
\nonumber
\end{equation}
\begin{equation}
\begin{array}{lll}
\Delta_1 = (v_1+v_2+v_3)p'-(u_2+u_3)p & \phantom{v_2p'}\qquad & \Delta_2 = (v_1+v_2)p'-u_2p \\
\Lambda_1 = v_3p'-u_3p & & \Lambda_2 = v_2p'-u_2p
\end{array}
\nonumber
\end{equation}
\begin{equation}
\begin{array}{lll}
\Delta_1 = (u_1+u_2)p-v_2p' & \phantom{(v_1+v_2+v_3)p'}\qquad & \Delta_2 = (v_1+v_3)p'-u_3p \\
\Lambda_1 = v_2p'-u_2p & & \Lambda_2 = v_3p'-u_3p.
\end{array}
\nonumber
\end{equation}

\section{Structures}

Here we present the structures summarized in the introduced element $\mathcal{S}$. These components are similar for all form factors except for $G_A^v$.

When the single quark in the initial nucleon is at the bottom position, we get the following list of structures for all remaining form factors except for $G_A^s$. Its results require the exchange of $\gamma_\mu \rightarrow \gamma_\mu\gamma_5$. We find
\pagebreak[0]
\begin{eqnarray}
\mathcal{S}_1 &=& \bar{N}(P')\gamma_{\alpha_4}N(P)
\Tr [\gamma_\mu\pslash\gamma_{\alpha_2}\Deltaslash_2\gamma_{\alpha_3}\ppslash\gamma_{\alpha_1}\Deltaslash_1](VV+AA) \nonumber \\
\mathcal{S}_2 &=& \bar{N}(P')\gamma_{\alpha_4}\gamma_5N(P)
\Tr [\gamma_\mu\pslash\gamma_{\alpha_2}\Deltaslash_2\gamma_{\alpha_3}\gamma_5\ppslash\gamma_{\alpha_1}\Deltaslash_1](-AV-VA) \nonumber \\
\mathcal{S}_3 &=& \bar{N}(P')\gamma^{\lambda'}\gamma_{\alpha_4}\gamma^{\lambda}N(P)
\Tr [\gamma_\mu i\sigma_{\lambda p}\gamma_{\alpha_2}\Deltaslash_2\gamma_{\alpha_3}i\sigma_{\lambda' p'}\gamma_{\alpha_1}\Deltaslash_1](-TT) \nonumber
\end{eqnarray}
\begin{eqnarray}
\mathcal{S}_1 &=& \bar{N}(P')\gamma_{\alpha_4}N(P)
\Tr [\gamma_\mu\pslash\gamma_{\alpha_3}\Deltaslash_2\gamma_{\alpha_2}\ppslash\gamma_{\alpha_1}\Deltaslash_1](VV+AA) \nonumber \\
\mathcal{S}_2 &=& \bar{N}(P')\gamma_{\alpha_4}\gamma_5N(P)
\Tr [\gamma_\mu\pslash\gamma_{\alpha_3}\gamma_5\Deltaslash_2\gamma_{\alpha_2}\ppslash\gamma_{\alpha_1}\Deltaslash_1](-AV-VA) \nonumber \\
\mathcal{S}_3 &=& \bar{N}(P')\gamma^{\lambda'}\gamma_{\alpha_4}\gamma^{\lambda}N(P)
\Tr [\gamma_\mu i\sigma_{\lambda p}\gamma_{\alpha_3}\Deltaslash_2\gamma_{\alpha_2}i\sigma_{\lambda' p'}\gamma_{\alpha_1}\Deltaslash_1](-TT) \nonumber
\end{eqnarray}
\begin{eqnarray}
\mathcal{S}_1 &=& \bar{N}(P')\gamma_{\alpha_4}\Deltaslash_2\gamma_{\alpha_2}N(P)
\Tr [\gamma_\mu\pslash\gamma_{\alpha_3}\ppslash\gamma_{\alpha_1}\Deltaslash_1](VV+AA) \nonumber \\
\mathcal{S}_2 &=& \bar{N}(P')\gamma_{\alpha_4}\gamma_5\Deltaslash_2\gamma_{\alpha_2}N(P)
\Tr [\gamma_\mu\pslash\gamma_{\alpha_3}\gamma_5\ppslash\gamma_{\alpha_1}\Deltaslash_1](-AV-VA) \nonumber \\
\mathcal{S}_3 &=& \bar{N}(P')\gamma^{\lambda'}\gamma_{\alpha_4}\Deltaslash_2\gamma_{\alpha_2}\gamma^{\lambda}N(P)
\Tr [\gamma_\mu i\sigma_{\lambda p}\gamma_{\alpha_3}i\sigma_{\lambda' p'}\gamma_{\alpha_1}\Deltaslash_1](-TT) \nonumber
\end{eqnarray}
\begin{eqnarray}
\mathcal{S}_1 &=& \bar{N}(P')\gamma_{\alpha_2}\Deltaslash_2\gamma_{\alpha_4}N(P)
\Tr [\gamma_\mu\pslash\gamma_{\alpha_3}\ppslash\gamma_{\alpha_1}\Deltaslash_1](VV+AA) \nonumber \\
\mathcal{S}_2 &=& \bar{N}(P')\gamma_{\alpha_2}\Deltaslash_2\gamma_{\alpha_4}\gamma_5N(P)
\Tr [\gamma_\mu\pslash\gamma_{\alpha_3}\gamma_5\ppslash\gamma_{\alpha_1}\Deltaslash_1](-AV-VA) \nonumber \\
\mathcal{S}_3 &=& \bar{N}(P')\gamma^{\lambda'}\gamma_{\alpha_2}\Deltaslash_2\gamma_{\alpha_4}\gamma^{\lambda}N(P)
\Tr [\gamma_\mu i\sigma_{\lambda p}\gamma_{\alpha_3}i\sigma_{\lambda' p'}\gamma_{\alpha_1}\Deltaslash_1](-TT) \nonumber
\end{eqnarray}
\begin{eqnarray}
\mathcal{S}_1 &=& \bar{N}(P')\gamma_{\alpha_4}N(P)
\Tr [\gamma_\mu\pslash\gamma_{\alpha_2}\ppslash\gamma_{\alpha_3}\Deltaslash_2\gamma_{\alpha_1}\Deltaslash_1](VV+AA) \nonumber \\
\mathcal{S}_2 &=& \bar{N}(P')\gamma_{\alpha_4}\gamma_5N(P)
\Tr [\gamma_\mu\pslash\gamma_{\alpha_2}\ppslash\gamma_{\alpha_3}\gamma_5\Deltaslash_2\gamma_{\alpha_1}\Deltaslash_1](-AV-VA) \nonumber \\
\mathcal{S}_3 &=& \bar{N}(P')\gamma^{\lambda'}\gamma_{\alpha_4}\gamma^{\lambda}N(P)
\Tr [\gamma_\mu i\sigma_{\lambda p}\gamma_{\alpha_2}i\sigma_{\lambda' p'}\gamma_{\alpha_3}\Deltaslash_2\gamma_{\alpha_1}\Deltaslash_1](-TT) \nonumber
\end{eqnarray}
\begin{eqnarray}
\mathcal{S}_1 &=& \bar{N}(P')\gamma_{\alpha_2}N(P)
\Tr [\gamma_\mu\pslash\gamma_{\alpha_4}\ppslash\gamma_{\alpha_3}\Deltaslash_2\gamma_{\alpha_1}\Deltaslash_1](VV+AA) \nonumber \\
\mathcal{S}_2 &=& \bar{N}(P')\gamma_{\alpha_2}\gamma_5N(P)
\Tr [\gamma_\mu\pslash\gamma_{\alpha_4}\ppslash\gamma_{\alpha_3}\Deltaslash_2\gamma_{\alpha_1}\gamma_5\Deltaslash_1](-AV-VA) \nonumber \\
\mathcal{S}_3 &=& \bar{N}(P')\gamma^{\lambda'}\gamma_{\alpha_2}\gamma^{\lambda}N(P)
\Tr [\gamma_\mu i\sigma_{\lambda p}\gamma_{\alpha_4}i\sigma_{\lambda' p'}\gamma_{\alpha_3}\Deltaslash_2\gamma_{\alpha_1}\Deltaslash_1](-TT) \nonumber
\end{eqnarray}
\begin{eqnarray}
\mathcal{S}_1 &=& \bar{N}(P')\gamma_{\alpha_4}N(P)
\Tr [\gamma_\mu\Deltaslash_1\gamma_{\alpha_1}\pslash\gamma_{\alpha_2}\ppslash\gamma_{\alpha_3}\Deltaslash_2](VV+AA) \nonumber \\
\mathcal{S}_2 &=& \bar{N}(P')\gamma_{\alpha_4}\gamma_5N(P)
\Tr [\gamma_\mu\Deltaslash_1\gamma_{\alpha_1}\pslash\gamma_{\alpha_2}\ppslash\gamma_{\alpha_3}\gamma_5\Deltaslash_2](-AV-VA) \nonumber \\
\mathcal{S}_3 &=& \bar{N}(P')\gamma^{\lambda'}\gamma_{\alpha_4}\gamma^{\lambda}N(P)
\Tr [\gamma_\mu\Deltaslash_1\gamma_{\alpha_1}i\sigma_{\lambda p}\gamma_{\alpha_2}i\sigma_{\lambda' p'}\gamma_{\alpha_3}\Deltaslash_2](-TT). \nonumber
\end{eqnarray}

When the considered quark is at the top position, we get again a list of structures for all remaining form factors except for $G_A^s$. Its results require the same exchange as applied in the previous case. We get
\pagebreak[0]
\begin{eqnarray}
\mathcal{S}_1 &=& \bar{N}(P')\gamma_{\alpha_1}\Deltaslash_1\gamma_\mu N(P)
\Tr [\pslash\gamma_{\alpha_2}\Deltaslash_2\gamma_{\alpha_3}\ppslash\gamma_{\alpha_4}](VV+AA) \nonumber \\
\mathcal{S}_2 &=& \bar{N}(P')\gamma_{\alpha_1}\gamma_5\Deltaslash_1\gamma_\mu N(P)
\Tr [\pslash\gamma_{\alpha_2}\gamma_5\Deltaslash_2\gamma_{\alpha_3}\ppslash\gamma_{\alpha_4}](-AV-VA) \nonumber \\
\mathcal{S}_3 &=& \bar{N}(P')\gamma^{\lambda'}\gamma_{\alpha_1}\Deltaslash_1\gamma_\mu\gamma^{\lambda}N(P)
\Tr [i\sigma_{\lambda p}\gamma_{\alpha_2}\Deltaslash_2\gamma_{\alpha _3}i\sigma_{\lambda' p'}\gamma_{\alpha_4}](-TT) \nonumber
\end{eqnarray}
\begin{eqnarray}
\mathcal{S}_1 &=& \bar{N}(P')\gamma_{\alpha_1}\Deltaslash_1\gamma_\mu N(P)
\Tr [\pslash\gamma_{\alpha_3}\Deltaslash_2\gamma_{\alpha_2}\ppslash\gamma_{\alpha_4}](VV+AA) \nonumber \\
\mathcal{S}_2 &=& \bar{N}(P')\gamma_{\alpha_1}\gamma_5\Deltaslash_1\gamma_\mu N(P)
\Tr [\pslash\gamma_{\alpha_3}\Deltaslash_2\gamma_{\alpha_2}\gamma_5\ppslash\gamma_{\alpha_4}](-AV-VA) \nonumber \\
\mathcal{S}_3 &=& \bar{N}(P')\gamma^{\lambda'}\gamma_{\alpha_1}\Deltaslash_1\gamma_\mu\gamma^{\lambda}N(P)
\Tr [i\sigma_{\lambda p}\gamma_{\alpha_3}\Deltaslash_2\gamma_{\alpha_2}i\sigma_{\lambda' p'}\gamma_{\alpha_4}](-TT) \nonumber
\end{eqnarray}
\begin{eqnarray}
\mathcal{S}_1 &=& \bar{N}(P')\gamma_{\alpha_1}\Deltaslash_1\gamma_\mu N(P)
\Tr [\pslash\gamma_{\alpha_3}\ppslash\gamma_{\alpha_4}\Deltaslash_2\gamma_{\alpha_2}](VV+AA) \nonumber \\
\mathcal{S}_2 &=& \bar{N}(P')\gamma_{\alpha_1}\gamma_5\Deltaslash_1\gamma_\mu N(P)
\Tr [\pslash\gamma_{\alpha_3}\ppslash\gamma_{\alpha_4}\Deltaslash_2\gamma_{\alpha_2}\gamma_5](-AV-VA) \nonumber \\
\mathcal{S}_3 &=& \bar{N}(P')\gamma^{\lambda'}\gamma_{\alpha_1}\Deltaslash_1\gamma_\mu\gamma^{\lambda}N(P)
\Tr [i\sigma_{\lambda p}\gamma_{\alpha_3}i\sigma_{\lambda' p'}\gamma_{\alpha_4}\Deltaslash_2\gamma_{\alpha_2}](-TT) \nonumber
\end{eqnarray}
\begin{eqnarray}
\mathcal{S}_1 &=& \bar{N}(P')\gamma_{\alpha_1}\Deltaslash_1\gamma_\mu N(P)
\Tr [\pslash\gamma_{\alpha_3}\ppslash\gamma_{\alpha_2}\Deltaslash_2\gamma_{\alpha_4}](VV+AA) \nonumber \\
\mathcal{S}_2 &=& \bar{N}(P')\gamma_{\alpha_1}\gamma_5\Deltaslash_1\gamma_\mu N(P)
\Tr [\pslash\gamma_{\alpha_3}\ppslash\gamma_{\alpha_2}\gamma_5\Deltaslash_2\gamma_{\alpha_4}](-AV-VA) \nonumber \\
\mathcal{S}_3 &=& \bar{N}(P')\gamma^{\lambda'}\gamma_{\alpha_1}\Deltaslash_1\gamma_\mu\gamma^{\lambda}N(P)
\Tr [i\sigma_{\lambda p}\gamma_{\alpha_3}i\sigma_{\lambda' p'}\gamma_{\alpha_2}\Deltaslash_2\gamma_{\alpha_4}](-TT) \nonumber
\end{eqnarray}
\begin{eqnarray}
\mathcal{S}_1 &=& \bar{N}(P')\gamma_{\alpha_3}\Deltaslash_2\gamma_{\alpha_1}\Deltaslash_1\gamma_\mu N(P)
\Tr [\pslash\gamma_{\alpha_2}\ppslash\gamma_{\alpha_4}](VV+AA) \nonumber \\
\mathcal{S}_2 &=& \bar{N}(P')\gamma_{\alpha_3}\Deltaslash_2\gamma_{\alpha_1}\gamma_5\Deltaslash_1\gamma_\mu N(P)
\Tr [\pslash\gamma_{\alpha_2}\gamma_5\ppslash\gamma_{\alpha_4}](-AV-VA) \nonumber \\
\mathcal{S}_3 &=& \bar{N}(P')\gamma ^{\lambda'}\gamma_{\alpha_3}\Deltaslash_2\gamma_{\alpha_1}\Deltaslash_1\gamma_\mu\gamma^{\lambda}N(P)
\Tr [i\sigma _{\lambda p}\gamma_{\alpha_2}i\sigma_{\lambda' p'}\gamma_{\alpha_4}](-TT) \nonumber
\end{eqnarray}
\begin{eqnarray}
\mathcal{S}_1 &=& \bar{N}(P')\gamma_{\alpha_3}\Deltaslash_2\gamma_{\alpha_1}\Deltaslash_1\gamma_\mu N(P)
\Tr [\pslash\gamma_{\alpha_4}\ppslash\gamma_{\alpha_2}](VV+AA) \nonumber \\
\mathcal{S}_2 &=& \bar{N}(P')\gamma_{\alpha_3}\Deltaslash_2\gamma_{\alpha_1}\gamma_5\Deltaslash_1\gamma_\mu N(P)
\Tr [\pslash\gamma_{\alpha_4}\ppslash\gamma_{\alpha_2}\gamma_5](-AV-VA) \nonumber \\
\mathcal{S}_3 &=& \bar{N}(P')\gamma^{\lambda'}\gamma_{\alpha_3}\Deltaslash_2\gamma_{\alpha_1}\Deltaslash_1\gamma_\mu\gamma^{\lambda}N(P)
\Tr [i\sigma_{\lambda p}\gamma_{\alpha_4}i\sigma_{\lambda' p'}\gamma_{\alpha_2}](-TT) \nonumber
\end{eqnarray}
\begin{eqnarray}
\mathcal{S}_1 &=& \bar{N}(P')\gamma_{\alpha_3}\Deltaslash_2\gamma_\mu\Deltaslash_1\gamma_{\alpha_1}N(P)
\Tr [\pslash\gamma_{\alpha_2}\ppslash\gamma_{\alpha_4}](VV+AA) \nonumber \\
\mathcal{S}_2 &=& \bar{N}(P')\gamma_{\alpha_3}\Deltaslash_2\gamma_\mu\Deltaslash_1\gamma_{\alpha_1}\gamma_5N(P)
\Tr [\pslash\gamma_{\alpha_2}\gamma_5\ppslash\gamma_{\alpha_4}](-AV-VA) \nonumber \\
\mathcal{S}_3 &=& \bar{N}(P')\gamma^{\lambda'}\gamma_{\alpha_3}\Deltaslash_2\gamma_\mu\Deltaslash_1\gamma_{\alpha_1}\gamma^{\lambda}N(P)
\Tr [i\sigma_{\lambda p}\gamma_{\alpha_2}i\sigma_{\lambda' p'}\gamma_{\alpha_4}](-TT). \nonumber
\end{eqnarray}

We obtain the following list of structures for $G_A^v$. The discussed quark must be at the bottom position here, so that
\pagebreak[0]
\begin{eqnarray}
\mathcal{S}_1 &=& \bar{N}(P')\gamma_{\alpha_3}\Deltaslash_2\gamma_{\alpha_2}\pslash\gamma_\mu\gamma_5\Deltaslash_1\gamma_{\alpha_1}\ppslash\gamma_{\alpha _4}N(P)
(VV+AA+AV+VA) \nonumber \\
\mathcal{S}_2 &=& \bar{N}(P')\gamma^{\lambda'}\gamma_{\alpha_3}\Deltaslash_2\gamma_{\alpha_2}\pslash\gamma_\mu\gamma_5\Deltaslash_1\gamma_{\alpha_1}
i\sigma_{\lambda' p'}\gamma_{\alpha_4}N(P)(-TV+TA) \nonumber \\
\mathcal{S}_3 &=& \bar{N}(P')\gamma_{\alpha_3}\Deltaslash_2\gamma_{\alpha_2}i\sigma_{\lambda p}\gamma_\mu\gamma_5
\Deltaslash_1\gamma_{\alpha_1}\ppslash\gamma_{\alpha_4}\gamma^{\lambda}N(P)(VT-AT) \nonumber \\
\mathcal{S}_4 &=& \bar{N}(P')\gamma^{\lambda'}\gamma_{\alpha_3}\Deltaslash_2\gamma_{\alpha_2}i\sigma_{\lambda p}\gamma_\mu\gamma_5
\Deltaslash_1\gamma_{\alpha_1}i\sigma_{\lambda' p'}\gamma_{\alpha_4}\gamma^{\lambda}N(P)(-TT) \nonumber
\end{eqnarray}
\begin{eqnarray}
\mathcal{S}_1 &=& \bar{N}(P')\gamma_{\alpha_2}\Deltaslash_2\gamma_{\alpha_3}\pslash\gamma_\mu\gamma_5\Deltaslash_1\gamma_{\alpha_1}\ppslash\gamma_{\alpha_4}N(P)
(VV+AA+AV+VA) \nonumber \\
\mathcal{S}_2 &=& \bar{N}(P')\gamma^{\lambda'}\gamma_{\alpha_2}\Deltaslash_2\gamma_{\alpha_3}\pslash\gamma_\mu\gamma_5\Deltaslash_1\gamma_{\alpha_1}
i\sigma_{\lambda' p'}\gamma_{\alpha_4}N(P)(-TV+TA) \nonumber \\
\mathcal{S}_3 &=& \bar{N}(P')\gamma_{\alpha_2}\Deltaslash_2\gamma_{\alpha_3}i\sigma_{\lambda p}\gamma_\mu\gamma_5
\Deltaslash_1\gamma_{\alpha_1}\ppslash\gamma_{\alpha_4}\gamma^{\lambda}N(P)(VT-AT) \nonumber \\
\mathcal{S}_4 &=& \bar{N}(P')\gamma^{\lambda'}\gamma_{\alpha_2}\Deltaslash_2\gamma_{\alpha_3}i\sigma_{\lambda p}\gamma_\mu\gamma_5
\Deltaslash_1\gamma_{\alpha_1}i\sigma_{\lambda' p'}\gamma_{\alpha_4}\gamma^{\lambda}N(P)(-TT) \nonumber
\end{eqnarray}
\begin{eqnarray}
\mathcal{S}_1 &=& \bar{N}(P')\gamma_{\alpha_3}\pslash\gamma_\mu\gamma_5\Deltaslash_1\gamma_{\alpha_1}\ppslash\gamma_{\alpha_4}\Deltaslash_2\gamma_{\alpha_2}N(P)
(VV+AA+AV+VA) \nonumber \\
\mathcal{S}_2 &=& \bar{N}(P')\gamma^{\lambda'}\gamma_{\alpha_3}\pslash\gamma_\mu\gamma_5\Deltaslash_1\gamma_{\alpha_1}
i\sigma_{\lambda' p'}\gamma_{\alpha_4}\Deltaslash_2\gamma_{\alpha_2}N(P)(-TV+TA) \nonumber \\
\mathcal{S}_3 &=& \bar{N}(P')\gamma_{\alpha_3}i\sigma_{\lambda p}\gamma_\mu\gamma_5
\Deltaslash_1\gamma_{\alpha_1}\ppslash\gamma_{\alpha_4}\Deltaslash_2\gamma_{\alpha_2}\gamma^{\lambda}N(P)(VT-AT) \nonumber \\
\mathcal{S}_4 &=& \bar{N}(P')\gamma^{\lambda'}\gamma_{\alpha_3}i\sigma_{\lambda p}\gamma_\mu\gamma_5\Deltaslash_1\gamma_{\alpha_1}
i\sigma_{\lambda' p'}\gamma_{\alpha_4}\Deltaslash_2\gamma_{\alpha_2}\gamma^{\lambda}N(P)(-TT) \nonumber
\end{eqnarray}
\begin{eqnarray}
\mathcal{S}_1 &=& \bar{N}(P')\gamma_{\alpha_1}\pslash\gamma_\mu\gamma_5\Deltaslash_1\gamma_{\alpha_3}\ppslash\gamma_{\alpha_4}\Deltaslash_2\gamma_{\alpha_2}N(P)
(VV+AA+AV+VA) \nonumber \\
\mathcal{S}_2 &=& \bar{N}(P')\gamma^{\lambda'}\gamma_{\alpha_1}\pslash\gamma_\mu\gamma_5\Deltaslash_1\gamma_{\alpha_3}
i\sigma_{\lambda' p'}\gamma_{\alpha_4}\Deltaslash_2\gamma_{\alpha_2}N(P)(-TV+TA) \nonumber \\
\mathcal{S}_3 &=& \bar{N}(P')\gamma_{\alpha_1}i\sigma_{\lambda p}\gamma_\mu\gamma_5
\Deltaslash_1\gamma_{\alpha_3}\ppslash\gamma_{\alpha_4}\Deltaslash_2\gamma_{\alpha_2}\gamma^{\lambda}N(P)(VT-AT) \nonumber \\
\mathcal{S}_4 &=& \bar{N}(P')\gamma^{\lambda'}\gamma_{\alpha_1}i\sigma_{\lambda p}\gamma_\mu\gamma_5\Deltaslash_1\gamma_{\alpha_3}
i\sigma_{\lambda' p'}\gamma_{\alpha_4}\Deltaslash_2\gamma_{\alpha_2}\gamma^{\lambda}N(P)(-TT) \nonumber
\end{eqnarray}
\begin{eqnarray}
\mathcal{S}_1 &=& \bar{N}(P')\gamma_{\alpha_2}\pslash\gamma_\mu\gamma_5\Deltaslash_1\gamma_{\alpha_1}\Deltaslash_2\gamma_{\alpha_3}\ppslash\gamma_{\alpha_4}N(P)
(VV+AA+AV+VA) \nonumber \\
\mathcal{S}_2 &=& \bar{N}(P')\gamma^{\lambda'}\gamma_{\alpha_2}\pslash\gamma_\mu\gamma_5
\Deltaslash_1\gamma_{\alpha_1}\Deltaslash_2\gamma_{\alpha_3}i\sigma_{\lambda' p'}\gamma_{\alpha_4}N(P)(-TV+TA) \nonumber \\
\mathcal{S}_3 &=& \bar{N}(P')\gamma_{\alpha_2}i\sigma_{\lambda p}\gamma_\mu\gamma_5
\Deltaslash_1\gamma_{\alpha_1}\Deltaslash_2\gamma_{\alpha_3}\ppslash\gamma_{\alpha_4}\gamma^{\lambda}N(P)(VT-AT) \nonumber \\
\mathcal{S}_4 &=& \bar{N}(P')\gamma^{\lambda'}\gamma_{\alpha_2}i\sigma_{\lambda p}\gamma_\mu\gamma_5
\Deltaslash_1\gamma_{\alpha_1}\Deltaslash_2\gamma_{\alpha_3}i\sigma_{\lambda' p'}\gamma_{\alpha_4}\gamma^{\lambda}N(P)(-TT) \nonumber
\end{eqnarray}
\begin{eqnarray}
\mathcal{S}_1 &=& \bar{N}(P')\gamma_{\alpha_4}\pslash\gamma_\mu\gamma_5\Deltaslash_1\gamma_{\alpha_1}\Deltaslash_2\gamma_{\alpha_3}\ppslash\gamma_{\alpha_2}N(P)
(VV+AA+AV+VA) \nonumber \\
\mathcal{S}_2 &=& \bar{N}(P')\gamma^{\lambda'}\gamma_{\alpha_4}\pslash\gamma_\mu\gamma_5
\Deltaslash_1\gamma_{\alpha_1}\Deltaslash_2\gamma_{\alpha_3}i\sigma_{\lambda' p'}\gamma_{\alpha_2}N(P)(-TV+TA) \nonumber \\
\mathcal{S}_3 &=& \bar{N}(P')\gamma_{\alpha_4}i\sigma_{\lambda p}\gamma_\mu\gamma_5
\Deltaslash_1\gamma_{\alpha_1}\Deltaslash_2\gamma_{\alpha_3}\ppslash\gamma_{\alpha_2}\gamma^{\lambda}N(P)(VT-AT) \nonumber \\
\mathcal{S}_4 &=& \bar{N}(P')\gamma^{\lambda'}\gamma_{\alpha_4}i\sigma_{\lambda p}\gamma_\mu\gamma_5
\Deltaslash_1\gamma_{\alpha_1}\Deltaslash_2\gamma_{\alpha_3}i\sigma_{\lambda' p'}\gamma_{\alpha_2}\gamma^{\lambda}N(P)(-TT) \nonumber
\end{eqnarray}
\begin{eqnarray}
\mathcal{S}_1 &=& \bar{N}(P')\gamma_{\alpha_2}\pslash\gamma_{\alpha_1}\Deltaslash_1\gamma_\mu\gamma_5\Deltaslash_2\gamma_{\alpha_3}\ppslash\gamma_{\alpha_4}N(P)
(VV+AA+AV+VA) \nonumber \\
\mathcal{S}_2 &=& \bar{N}(P')\gamma^{\lambda'}\gamma_{\alpha_2}\pslash\gamma_{\alpha_1}\Deltaslash_1\gamma_\mu\gamma_5
\Deltaslash_2\gamma_{\alpha_3}i\sigma_{\lambda' p'}\gamma_{\alpha_4}N(P)(-TV+TA) \nonumber \\
\mathcal{S}_3 &=& \bar{N}(P')\gamma_{\alpha_2}i\sigma_{\lambda p}\gamma_{\alpha_1}\Deltaslash_1\gamma_\mu\gamma_5
\Deltaslash_2\gamma_{\alpha_3}\ppslash\gamma_{\alpha_4}\gamma^{\lambda}N(P)(VT-AT) \nonumber \\
\mathcal{S}_4 &=& \bar{N}(P')\gamma^{\lambda'}\gamma_{\alpha_2}i\sigma_{\lambda p}\gamma_{\alpha_1}\Deltaslash_1\gamma_\mu\gamma_5
\Deltaslash_2\gamma_{\alpha_3}i\sigma_{\lambda' p'}\gamma_{\alpha_4}\gamma^{\lambda}N(P)(-TT). \nonumber
\end{eqnarray}

\section{Results}

We start with the results for $G_M^p$ and mention that the results for $G_M^n$ can be obtained by the interchange of $e_u \leftrightarrow e_d$. Furthermore, the results for $G_A^s$ differ from them by the absence of the quark charge, the exchange of $\gamma_\mu \rightarrow \gamma_\mu\gamma_5$, and an opposite overall sign in the results of the seventh diagram in both considered quark position cases.

First, we consider the case when the discussed quark is at the bottom position. In this case, we obtain
\pagebreak[0]
\begin{equation}
\frac{(4\pi\bar{\alpha}_s)^2}{216}\frac{e_u}{Q^4}\int\frac{[\intd\!u]}{u_3(u_2+u_3)^2}\frac{[\intd\!v]}{v_3(v_2+v_3)^2}\left[(V-A)^2+4T^2\right]\bar{N}(P')\gamma_\mu N(P)
\nonumber
\end{equation}
\begin{equation}
0
\nonumber
\end{equation}
\begin{equation}
\frac{(4\pi\bar{\alpha}_s)^2}{216}\frac{e_u}{Q^4}\int\frac{[\intd\!u]}{u_2(u_2+u_3)^2}\frac{[\intd\!v]}{v_2(v_2+v_3)^2}\left[(V-A)^2+4T^2\right]\bar{N}(P')\gamma_\mu N(P)
\nonumber
\end{equation}
\begin{equation}
0
\nonumber
\end{equation}
\begin{equation}
\frac{(4\pi\bar{\alpha}_s)^2}{216}\frac{e_u}{Q^4}\int\frac{[\intd\!u]}{u_2u_3(u_2+u_3)}\frac{[\intd\!v]}{v_2v_3(v_1+v_3)}\left[-4T^2\right]\bar{N}(P')\gamma_\mu N(P)
\nonumber
\end{equation}
\begin{equation}
\frac{(4\pi\bar{\alpha}_s)^2}{216}\frac{e_u}{Q^4}\int\frac{[\intd\!u]}{u_2u_3(u_2+u_3)}\frac{[\intd\!v]}{v_2v_3(v_1+v_2)}\left[-(V-A)^2\right]\bar{N}(P')\gamma_\mu N(P)
\nonumber
\end{equation}
\begin{equation}
\frac{(4\pi\bar{\alpha}_s)^2}{216}\frac{e_u}{Q^4}\int\frac{[\intd\!u]}{u_2u_3(u_1+u_2)}\frac{[\intd\!v]}{v_2v_3(v_1+v_3)}\left[(V+A)^2\right]\bar{N}(P')\gamma_\mu N(P).
\nonumber
\end{equation}

Second, we consider the case when the discussed quark is at the top position. In this case, we obtain
\pagebreak[0]
\begin{equation}
\frac{(4\pi\bar{\alpha}_s)^2}{216}\frac{e_d}{Q^4}\int\frac{[\intd\!u]}{u_1(u_1+u_2)^2}\frac{[\intd\!v]}{v_1(v_1+v_2)^2}\left[2(V^2+A^2)\right]\bar{N}(P')\gamma_\mu N(P)
\nonumber
\end{equation}
\begin{equation}
0
\nonumber
\end{equation}
\begin{equation}
\frac{(4\pi\bar{\alpha}_s)^2}{216}\frac{e_d}{Q^4}\int\frac{[\intd\!u]}{u_2(u_1+u_2)^2}\frac{[\intd\!v]}{v_2(v_1+v_2)^2}\left[2(V^2+A^2)\right]\bar{N}(P')\gamma_\mu N(P)
\nonumber
\end{equation}
\begin{equation}
0
\nonumber
\end{equation}
\begin{equation}
\frac{(4\pi\bar{\alpha}_s)^2}{216}\frac{e_d}{Q^4}\int\frac{[\intd\!u]}{u_1u_2(u_1+u_2)}\frac{[\intd\!v]}{v_1v_2(v_1+v_3)}\left[-(V+A)^2\right]\bar{N}(P')\gamma_\mu N(P)
\nonumber
\end{equation}
\begin{equation}
\frac{(4\pi\bar{\alpha}_s)^2}{216}\frac{e_d}{Q^4}\int\frac{[\intd\!u]}{u_1u_2(u_1+u_2)}\frac{[\intd\!v]}{v_1v_2(v_2+v_3)}\left[-(V-A)^2\right]\bar{N}(P')\gamma_\mu N(P)
\nonumber
\end{equation}
\begin{equation}
\frac{(4\pi\bar{\alpha}_s)^2}{216}\frac{e_d}{Q^4}\int\frac{[\intd\!u]}{u_1u_2(u_2+u_3)}\frac{[\intd\!v]}{v_1v_2(v_1+v_3)}\left[4T^2\right]\bar{N}(P')\gamma_\mu N(P).
\nonumber
\end{equation}

Finally, we present the results for $G_A^v$. The discussed quark must be at the bottom position here, so that
\pagebreak[0]
\begin{equation}
\frac{(4\pi\bar{\alpha}_s)^2}{216}\frac{1}{Q^4}\int\frac{[\intd\!u]}{u_3(u_2+u_3)^2}\frac{[\intd\!v]}{v_2(v_2+v_3)^2}\left[2(T(V-A)+(V-A)T)\right]
\bar{N}(P')\gamma_\mu\gamma_5N(P)
\nonumber
\end{equation}
\begin{equation}
0
\nonumber
\end{equation}
\begin{equation}
\frac{(4\pi\bar{\alpha}_s)^2}{216}\frac{1}{Q^4}\int\frac{[\intd\!u]}{u_2(u_2+u_3)^2}\frac{[\intd\!v]}{v_3(v_2+v_3)^2}\left[2(T(V-A)+(V-A)T)\right]
\bar{N}(P')\gamma_\mu\gamma_5N(P)
\nonumber
\end{equation}
\begin{equation}
0
\nonumber
\end{equation}
\begin{equation}
\frac{(4\pi\bar{\alpha}_s)^2}{216}\frac{1}{Q^4}\int\frac{[\intd\!u]}{u_2u_3(u_2+u_3)}\frac{[\intd\!v]}{v_2v_3(v_1+v_2)}\left[2(A-V)T\right]
\bar{N}(P')\gamma_\mu\gamma_5N(P)
\nonumber
\end{equation}
\begin{equation}
\frac{(4\pi\bar{\alpha}_s)^2}{216}\frac{1}{Q^4}\int\frac{[\intd\!u]}{u_2u_3(u_2+u_3)}\frac{[\intd\!v]}{v_2v_3(v_1+v_3)}\left[2T(A-V)\right]
\bar{N}(P')\gamma_\mu\gamma_5N(P)
\nonumber
\end{equation}
\begin{equation}
\frac{(4\pi\bar{\alpha}_s)^2}{216}\frac{1}{Q^4}\int\frac{[\intd\!u]}{u_2u_3(u_1+u_2)}\frac{[\intd\!v]}{v_2v_3(v_1+v_2)}\left[(V+A)(V+A)\right]
\bar{N}(P')\gamma_\mu\gamma_5N(P).
\nonumber
\end{equation}

$\phantom{X}$

We want to thank Dr. Dr. N. G. Stefanis for improving this work with substantial discussions and useful critical remarks. Moreover, we thank Prof. Dr. M. V. Polyakov for enabling this work. We also want to thank Prof. Dr. A. Metz and F. Grimme for showing us the discussed discrepancy in previous works, which was a main issue in this work.

The work has been supported by BMBF grant 06BO9012.

\bibliographystyle{h-physrev3}
\bibliography{Literature}

\begin{thebibliography}{10}

\bibitem{Lepage:1979a}
G.~P. {Lepage} and S.~J. {Brodsky},
\newblock Phys. Lett. {\bf B 87}, 359 (1979).

\bibitem{Lepage:1979b}
G.~P. {Lepage} and S.~J. {Brodsky},
\newblock Phys. Rev. Lett. {\bf 43}, 545 (1979).

\bibitem{Lepage:1980}
G.~P. {Lepage} and S.~J. {Brodsky},
\newblock Phys. Rev. {\bf D 22}, 2157 (1980).

\bibitem{Efremov:1980a}
A.~V. {Efremov} and A.~V. {Radyushkin},
\newblock Phys. Lett. {\bf B 94}, 245 (1980).

\bibitem{Efremov:1980b}
A.~V. {Efremov} and A.~V. {Radyushkin},
\newblock Theor. Math. Phys. {\bf 42}, 97 (1980).

\bibitem{Efremov:1981}
A.~V. {Efremov} and A.~V. {Radyushkin},
\newblock Theor. Math. Phys. {\bf 44}, 774 (1981).

\bibitem{Carlson:1987a}
C.~E. {Carlson} and F.~{Gross},
\newblock Phys. Rev. {\bf D 36}, 2060 (1987).

\bibitem{Carlson:1986}
C.~E. {Carlson} and J.~L. {Poor},
\newblock Phys. Rev. {\bf D 34}, 1478 (1986).

\bibitem{Carlson:1987c}
C.~E. {Carlson}, M.~{Gari}, and N.~G. {Stefanis},
\newblock Phys. Rev. Lett. {\bf 58}, 1308 (1987).

\bibitem{Carlson:1987b}
C.~E. {Carlson} and J.~L. {Poor},
\newblock Phys. Rev. {\bf D 36}, 2169 (1987).

\bibitem{Chernyak:1984a}
V.~L. {Chernyak} and I.~R. {Zhitnitsky},
\newblock Nucl. Phys. {\bf B 246}, 52 (1984).

\bibitem{Chernyak:1984b}
V.~L. {Chernyak} and A.~R. {Zhitnitsky},
\newblock Phys. Rep. {\bf 112}, 173 (1984).

\bibitem{Gari:1986}
M.~{Gari} and N.~G. {Stefanis},
\newblock Phys. Lett. {\bf B 175}, 462 (1986).

\bibitem{Stefanis:1989}
N.~G. {Stefanis},
\newblock Phys. Rev. {\bf D 40}, 2305 (1989).

\bibitem{Chernyak:1989}
V.~L. {Chernyak}, A.~A. {Ogloblin}, and I.~R. {Zhitnitsky},
\newblock Z. Phys. {\bf C 42}, 583 (1989).

\bibitem{Ji:1987}
C.~R. {Ji}, A.~F. {Sill}, and R.~M. {Lombard},
\newblock Phys. Rev. {\bf D 36}, 165 (1987).

\bibitem{Isgur:1989}
N.~{Isgur} and C.~H.~L. {Smith},
\newblock Nucl. Phys. {\bf B 317}, 526 (1989).

\bibitem{Botts:1989}
J.~{Botts} and G.~F. {Sterman},
\newblock Nucl. Phys. {\bf B 325}, 62 (1989).

\bibitem{Huang:1991}
T.~{Huang}, Q.~{Shen}, and P.~{Kroll},
\newblock Z. Phys. {\bf C 50}, 139 (1991).

\bibitem{Li:1992}
H.~{Li} and G.~F. {Sterman},
\newblock Nucl. Phys. {\bf B 381}, 129 (1992).

\bibitem{Li:1993}
H.~{Li},
\newblock Phys. Rev. {\bf D 48}, 4243 (1993).

\bibitem{Bolz:1995a}
J.~{Bolz}, R.~{Jakob}, P.~{Kroll}, M.~{Bergmann}, and N.~G. {Stefanis},
\newblock Z. Phys. {\bf C 66}, 267 (1995).

\bibitem{Bolz:1995b}
J.~{Bolz}, R.~{Jakob}, P.~{Kroll}, M.~{Bergmann}, and N.~G. {Stefanis},
\newblock Phys. Lett. {\bf B 342}, 345 (1995).

\bibitem{Stefanis:1999}
N.~G. {Stefanis},
\newblock Eur. Phys. J. direct {\bf C 7}, 1 (1999).

\bibitem{Brooks:2000}
T.~C. {Brooks} and L.~J. {Dixon},
\newblock Phys. Rev. {\bf D 62}, 114021 (2000).

\bibitem{Thomson:2006}
R.~{Thomson}, A.~{Pang}, and C.~R. {Ji},
\newblock Phys. Rev. {\bf D 73}, 054023 (2006).

\bibitem{Braun:2000}
V.~{Braun}, R.~J. {Fries}, N.~{Mahnke}, and E.~{Stein},
\newblock Nucl. Phys. {\bf B 589}, 381 (2000).

\bibitem{Braun:2006}
V.~M. {Braun}, A.~{Lenz}, and M.~{Wittmann},
\newblock Phys. Rev. {\bf D 73}, 094019 (2006).

\bibitem{Braun:2009}
V.~M. {Braun} {\em et~al.},
\newblock Phys. Rev. {\bf D 79}, 034504 (2009).

\bibitem{King:1987}
I.~D. {King} and C.~T. {Sachrajda},
\newblock Nucl. Phys. {\bf B 279}, 785 (1987).

\bibitem{Stefanis:1993}
N.~G. {Stefanis} and M.~{Bergmann},
\newblock Phys. Rev. {\bf D 47}, 3685 (1993).

\bibitem{Arnold:1986}
R.~G. {Arnold} {\em et~al.},
\newblock Phys. Rev. Lett. {\bf 57}, 174 (1986).

\bibitem{Sill:1993}
A.~F. {Sill} {\em et~al.},
\newblock Phys. Rev. {\bf D 48}, 29 (1993).

\bibitem{Rock:1982}
S.~{Rock} {\em et~al.},
\newblock Phys. Rev. Lett. {\bf 49}, 1139 (1982).

\bibitem{Rock:1992}
S.~{Rock} {\em et~al.},
\newblock Phys. Rev. {\bf D 46}, 24 (1992).

\bibitem{Kitagaki:1983}
T.~{Kitagaki} {\em et~al.},
\newblock Phys. Rev. {\bf D 28}, 436 (1983).

\end{thebibliography}

\end{document}